\documentstyle[12pt]{article}
\begin{document}

\begin{flushright}
TTP97-24\\ hep-ph/9710307\\ October 1997
\end{flushright}

\vspace{9mm}

\begin{center}

{\Large \bf \boldmath
$|V_{cb}|$ from semileptonic $B$ decays with 
two-loop QCD accuracy\footnote{Talk given at 5th
International Conference {\em Beyond the Standard Model},
Balholm, Norway, April 29 -- May 4, 1997.}}

\vspace{9mm}

{\large\bf Andrzej Czarnecki\footnote{Present address:
Physics Department,
Brookhaven National Laboratory,  Upton, New York 11973.} 
and Kirill Melnikov}

\vspace{2mm}

{\it Institut f\"ur Theoretische Teilchenphysik, \\
Universit\"at Karlsruhe,
D-76128 Karlsruhe, Germany}
\end{center}

\vspace{19mm}

\begin{abstract}
Recent progress in calculating ${\cal O}(\alpha _s ^2)$ corrections
to inclusive and exclusive semileptonic $b \to c$ transitions
is  reviewed. The impact of these corrections on the $|V_{cb}|$\
determination from both inclusive  and exclusive $B$ decays is
discussed.
\end{abstract}

\section{Introduction}
Semileptonic decays of $B$ mesons provide an opportunity to measure
the Cabibbo-Kobayashi-Maskawa (CKM) 
matrix parameter $|V_{cb}|$ with minimal
theoretical uncertainties (for a recent review, see
e.g. \cite{rev}). Its knowledge is important for testing
the unitarity of the CKM matrix and for better understanding of the
origin of CP violation.

The value recommended by the Particle Data Group last year is
$|V_{cb}| = 0.036~{\rm to}~0.046$ \cite{PDG}.  A lot of efforts have
since been made to reduce this range.  A recent summary of the
experimental results \cite{DiCiaccio} gives $|V_{cb}| = 0.0382\pm 
0.0019_{\rm exp}$ and the theoretical uncertainty is of a similar
size as the experimental error.  This improvement has been made
possible in part by calculations of the second order ($\sim
\alpha_s^2$) QCD corrections to the $b\to c$ transitions
\cite{zerorecoil,zerorecoila,CzMeUr,maxrecoil,techniques}.  In this
talk these results are briefly reviewed.

There are two main methods of the experimental determination of $|V_{cb}|$,
usually referred to as exclusive and inclusive.  In the former the end
point $\omega = 1$ of the differential distribution d$\Gamma/$d$\omega$
of the decay $B\to D^* l\nu$ ($\omega$ being the product of the $B$ and
$D^*$ four-velocities) is measured where the maximal information
 from heavy quark symmetry can be used \cite {HQET}.  The inclusive
method relies on the determination of the total semileptonic width in
the decays of $B$ mesons into charmed hadrons.  In order to extract
$|V_{cb}|$
 from the data the perturbative corrections must be accounted
for. Below we review the recent progress in their evaluation and
present the resulting  
values of $|V_{cb}|$.
 
Much of the recent progress in the heavy quark physics is due to the
Heavy Quark Expansion. In that framework
the nonperturbative effects are described using the operator product
expansion (OPE).  A correct treatment of perturbative corrections in a
manner consistent with the OPE is an interesting theoretical problem.
In the case of $b \to c$ transitions it is of practical importance for
a consistent extraction of the $|V_{cb}|$ from 
both inclusive and
exclusive $B$ decays.

Also, we mention that neither two-photon nor two-gluon corrections to
fermion decays had been  known until recently.  Such effects are
important not only for the $b\to c$ transitions but also for $b\to u$
and for the muon decay.

\section{Exclusive method}
The differential decay  rate for $B \to D^* l\nu _l$
can be written as
\begin{eqnarray}
\frac {{\rm d}\Gamma \left(B \to D^*l\nu_l\right)}{{\rm d}\omega} =
f(m_B,m_D^*,\omega)|V_{cb}|^2\;{\cal F}_{B \to D^*} ^2(\omega),
\end{eqnarray}
where $f$ is a known function (see e.g. \cite{HQET}) 
of $\omega$ and measurable particle masses. 
${\cal F}_{B \to D^*}$ is the form factor describing the transition
between the two 
hadronic states; its theoretical determination is most reliable at the
zero recoil point $\omega=1$. 
At that kinematical point the zero recoil 
sum rules for heavy flavor transitions \cite {optical1} predict the
value of ${\cal F}_{B \to D^*}$:
$$
|{\cal F}_{B \to D^*}| = \left [
\xi _A(\mu) - \frac {\xi _{\pi}(\mu)}{m_c^2} \mu _{\pi}^2(\mu)
-\frac {\xi_G(\mu)}{m_c^2} \mu _G^2(\mu)- 
\sum \limits _{\epsilon < \mu} |{\cal F}_{\rm exc}|^2
\right ]^{1/2} + {\cal O}\left( 1/m^3 \right).
$$
The coefficient functions $\xi _A,\;\xi _\pi$ and $\xi _G$ 
can be calculated perturbatively. On the other hand,
$
\mu _{\pi} ^2 =\langle B|\bar b \left( i \vec {D} \right ) ^2 b
|B\rangle/2M_B 
$
and 
$
\mu _G ^2 = - \langle B|\bar b \left(\vec \sigma \vec B \right)
b |B\rangle/2M_B
$
are  non--perturbative matrix elements of the kinetic and
chromomagnetic operators which appear in the OPE in the order
$\Lambda _{\rm QCD} ^2 /m_Q^2$.

The structure and the order of magnitude of the non--perturbative
contributions to the form factor have been understood since a long
time \cite{optical1}; the accuracy of the theoretical predictions
was to a large extent 
limited by unknown perturbative effects.
In order to find their magnitude
a complete calculation of ${\cal O}(\alpha _s ^2)$
corrections to $\xi _A(\mu)$ was undertaken.

To find $\xi _A(\mu)$ with 
${\cal O}(\alpha _s ^2)$ accuracy
one should calculate the ${\cal O}(\alpha _s ^2)$
corrections to the elastic $b \to c$ transition at zero recoil,
inelastic contribution with additional gluons in the final state,
the ${\cal O}(\alpha _s)$ correction to the Wilson 
coefficient $\xi _{\pi}$ of the kinetic operator in the sum rules
and also the two--loop power mixing of the kinetic operator with
the unit operator \cite{CzMeUr}.

The most technically demanding part of these calculations is the form
factor $\eta _A$ which describes the perturbative renormalization of
the axial current in the elastic $b \to c$ transition at the zero
recoil point to ${\cal O}(\alpha _s ^2)$ accuracy.  This problem has
been solved by two different methods.  First, in
Ref. \cite{zerorecoil}, the result was obtained by expanding the
Feynman diagrams in powers of $(m_b-m_c)/m_b$.  Later, in
Ref. \cite{zerorecoila} an exact analytical expression for $\eta _A$
has been obtained, confirming the results of \cite{zerorecoil}.

Other corrections necessary to evaluate $\xi _A(\mu)$ have been
recently calculated in Ref. \cite{CzMeUr}.  In an analysis of these
results it is convenient to separate the ${\cal O}(\alpha _s ^2)$
corrections into the part proportional to $\beta _0 \alpha _s^2$ (so
called BLM corrections \cite {blm}, first calculated for the $\eta_A$
function in \cite{Neub341}) and genuine ${\cal O}(\alpha _s ^2)$
corrections.  The complete resummation of the BLM corrections for $\xi
_A(\mu)$ was performed in \cite {blmope}.

The magnitude of the genuine ${\cal O}(\alpha _s ^2)$ corrections is
an indicator of our ability to control the higher order effects,
because, in contrast to BLM corrections, it is difficult to go beyond
the second order of perturbation theory at present.  The size of the
last calculated term can be used to estimate the uncalculated higher
order corrections.  In \cite {CzMeUr} we demonstrated that for any
reasonable choice of the scale parameter $\mu$ in the sum rules,
consistent with the $1/m_c$ expansion, the ${\cal O}(\alpha _s ^2)$
non-BLM corrections contribute at the level of less than $0.5 \%$ to
the coefficient function $\xi _A(\mu)$.

The second order non-BLM QCD corrections have been used to estimate
the uncertainty due to the remaining uncalculated higher order
effects.  A recent study  \cite{rev} gives
$$
{\cal F}_{B \to D^*} \simeq  0.91 \;-\;0.013\,
\frac{\mu_\pi^2-0.5\,\mbox{GeV}^2}{0.1\,\mbox{GeV}^2}\;\pm\;
0.02_{\rm excit}\;\pm\;0.01_{\rm pert}\;\pm\;0.025_{1/m^3},
$$
which is finally quoted as
${\cal F}_{B \to D^*} =  0.91\pm 0.06$.

With the recent experimental value \cite{DiCiaccio}
$
{\cal F}_{B \to D^*}(1)|V_{cb}| = (34.3\pm 1.6)\cdot 10^{-3}
$
we find
\begin{equation}
|V_{cb}| =  (37.7 \pm 1.8_{\rm exp} \pm 2.5_{\rm theor})\cdot 10^{-3}.
\end{equation}
A more optimistic estimate of the theoretical errors in ${\cal F}_{B
\to D^*}$ is given in \cite{NeubInt}; it leads to the theoretical
uncertainty in $|V_{cb}|$\ of $\pm 1.2\cdot 10^{-3}$ (see, however, a
discussion in \cite {rev}).

\section {Inclusive method}

In this method $|V_{cb}|$\ is determined from the inclusive semileptonic
decay width of the $B$ meson, $\Gamma _{\rm sl} (B \to X_c l\nu_l)$.
Applying the OPE to the decay width one finds that the
non--perturbative corrections in this case are suppressed by at least
two powers of the $b$--quark mass. Their magnitude is estimated to be
of the order of 5\% \cite {first,rev}.

With  the non--perturbative corrections under control, the size 
of the perturbative corrections was subject of some 
discussions in the literature. The necessity of performing a
complete calculation of the ${\cal O}(\alpha _s ^2)$ corrections
to the semileptonic decay width of the $b$ quark
has been repeatedly emphasized in recent years.

Computing complete ${\cal O}(\alpha _s ^2)$ correction to 
$\Gamma _{\rm sl}$ remains a daunting task
at present. In comparison with zero recoil calculations described in
the previous section, the main difficulties are: an additional
kinematical variable describing the invariant mass of the leptons 
and the presence of real radiation of one and two gluons.
In view of these difficulties it is necessary to find a 
way of estimating the ${\cal O}(\alpha _s ^2)$
effects with sufficient accuracy. 

To illustrate our idea we write the
semileptonic decay width  $\Gamma(b \to c l \nu _l)$ as
$
\Gamma _{\rm sl} = \int \limits _{0}^{(m_b-m_c)^2} {\rm d}q^2 
\left ({\rm d}\Gamma _{\rm sl}/{\rm d}q^2 \right ),
%\label {gammasl}
$
where $q^2$ is the invariant mass of the lepton pair.  The upper
boundary of this integral corresponds to the zero recoil point, where
the ${\cal O}(\alpha _s ^2)$ QCD correction is already known
\cite{zerorecoil,zerorecoila}.  If we compute the correction at the
other boundary ($q^2=0$) we might be able to estimate the total
correction.  More precisely, what we really need is a deviation of the
corrections from the BLM result, which is known for all $q^2$
\cite{inclblm,bbbsl}.

The technical details of the calculations at $q^2=0$, also called the
point of maximal recoil, are explained in detail in Refs. \cite
{maxrecoil,techniques}.  Here we sketch the main ideas of that
calculation.  Since an exact calculation is not yet feasible, we
construct an expansion in the velocity of the final quark. If the
final ($c$) quark is not much lighter than the $b$ quark this
expansion converges fairly well.  The expansion parameter is $\delta
\equiv (m_b-m_c)/m_b$.  Both the Feynman
amplitudes and the phase space can be represented as series in powers
and logarithms of $\delta$.

In case of  two--loop virtual corrections as well as 
in the emission  of two real
gluons, the expansion in $\delta$ is a Taylor expansion.
The situation is different in the case of the single gluon
radiation in diagrams where there is in addition one virtual gluon
loop.  Due to on--shell singularities of the one--loop diagrams,
the Taylor expansion  breaks down; in order to deal with such
diagrams we employ  the recently developed
method of ``eikonal expansions'' \cite{eikonal1}.

Let us now summarize the numerical results of these calculations 
\cite {maxrecoil}.  For the purpose of this discussion we use the mass
ratio of the $c$ and $b$ quarks $m_c/m_b = 0.3$.
It is again useful to separate the BLM and non--BLM corrections.
The complete resummation of the leading BLM corrections to
$\Gamma _{\rm sl}$ was performed in \cite {bbbsl}.

The ${\cal O}(\alpha _s ^2)$  non--BLM corrections 
calculated in \cite {maxrecoil} turn out to be quite small.
At $q_{\rm max}^2 = (m_b-m_c)^2$ (zero recoil) the 
non--BLM contributions give a $-0.1 \%$
correction to ${\rm d}\Gamma _{\rm sl}/{\rm d}q^2$.
At the maximal recoil
boundary $q^2 = 0$ the non--BLM correction 
constitutes about $1\%$.

There are good reasons to assume that the magnitude of the non--BLM
corrections increases with the increase in the phase space available
for the real radiation \cite{maxrecoil}. Based on this observation we
expect that the largest discrepancy between the BLM prediction and the
complete correction occurs at the lower end of the $q^2$ distribution,
i.e. for $q^2 = 0$.

Therefore, taking the value of the non--BLM corrections
at $q^2 = 0$ as an upper bound, we conclude
that  the non--BLM piece of the ${\cal O}(\alpha _s^2)$ correction
does not exceed the value of $1\%$ for any $q^2$.

We note that the accuracy of the $|V_{cb}|$ determination
 from the inclusive decays was sometimes questioned
because of  the unknown higher orders  corrections. 
The uncertainty related to perturbative effects and to the
quark mass values was thought to be as large as $10 \%$
in $|V_{cb}|$.
Our results show that the non--BLM effects are unlikely to cause such
large effects.

As for the BLM effects, considered in \cite {bbbsl}, we note that
the actual impact of all--orders BLM corrections and the values of
quark masses which should be used in the calculations of $\Gamma _{\rm
sl}$ should be considered simultaneously, because the uncertainties in
both effects tend to compensate each other (see \cite {rev} for a
recent discussion). A thoughtful choice of the normalization scale
for the quark masses \cite{five} significantly reduces higher order
perturbative corrections and the related uncertainty.

We quote finally the value of $|V_{cb}|$ obtained from inclusive 
measurements:
\begin{eqnarray}
|V_{cb}|\;\left (\Upsilon (4S)\right ) &=& \left(40.6 \pm 1.2_{\rm
exp} \right ) 
\left (1 \pm 0.05_{\rm th} \right)\cdot 10^{-3},
\\ \nonumber 
|V_{cb}|\;\left (Z \right ) &=& \left(43.1 \pm 0.6_{\rm exp} \right )
\left (1 \pm 0.05_{\rm th} \right)\cdot 10^{-3}.
\end{eqnarray}
To derive this value, we used experimental results as discussed
recently in \cite {DiCiaccio}. We give separately 
two values of $|V_{cb}|$, as obtained from the measurements at the $Z$
and at the $\Upsilon (4S)$ resonances.

For the theoretical uncertainties, we used the estimate 
presented in \cite {rev}, Eq. (8.5),
where we added all sources of the theoretical uncertainties linearly.
The theoretical uncertainty attributed 
in \cite {rev} to uncalculated higher order perturbative 
effects is consistent with our estimates of the ${\cal O}(\alpha _s ^2)$
corrections.

\section{Conclusions}
In this talk we have summarized the recent results
on ${\cal O}(\alpha _s ^2)$ corrections to $b \to c$ transition.
At the moment, complete ${\cal O}(\alpha _s ^2)$ corrections 
are known for the exclusive method and a reliable estimate exists for
the inclusive method for $|V_{cb}|$ determination
 from semileptonic $B$ decays.

Explicit calculations show that the ${\cal O}(\alpha _s ^2)$
corrections for both inclusive and exclusive transitions are dominated
by BLM corrections.  The genuine two-loop corrections contribute less
than $1\%$ to the differential decay rates.  With BLM effects resummed
for both inclusive and exclusive decays, this implies that no
significant uncertainty in $|V_{cb}|$ determination should be
attributed to uncalculated higher order effects.

Thus, an important  obstacle in the theoretical predictions for
the $|V_{cb}|$ determination 
 from semileptonic $B$ decays has been eliminated. In this
situation further improvement of theoretical predictions
seems only possible with more accurate estimates of the
non--perturbative effects (in particular in exclusive decays) and
more precise determination of the input parameters, such as quark
masses and numerical values of the non--perturbative matrix elements.

\section*{Acknowledgments} We thank Per Osland for the opportunity to
attend the conference ``Beyond the Standard Model V'' in Balholm,
Norway, and for partial support. We are grateful to Nikolai Uraltsev
for a fruitful collaboration.  This work was supported by BMBF under
grant number BMBF-057KA92P, and by Graduiertenkolleg
``Teilchenphysik'' at the University of Karlsruhe.

\end{document}